\documentclass[a4paper,12pt]{article}

\usepackage{amsthm}
\usepackage{amsmath,amssymb,latexsym,amsfonts,mathrsfs}
\usepackage[dvipdfmx]{graphicx}

\usepackage{fancyhdr}
\usepackage[top=30truemm, bottom=30truemm, left=25truemm, right=25truemm]{geometry}

\newcommand{\ep}{\varepsilon}
\newcommand{\nn}{\nonumber}

\newcommand{\SCR}[1]{{\mathscr #1}}

\newcommand{\CAL}[1]{{\cal #1}
}

\newcommand{\MAT}[1]{\left(\begin{array}{cccccccccc}#1\end{array}\right)}

\newcommand{\J}[1]{\left\langle #1 \right\rangle}

\newcommand{\CR}[1]{ {#1} }

\theoremstyle{definition}
\newtheorem{thm}{{\bf Theorem}}[section]

\newtheorem{lem}[thm]{{\bf Lemma}}
\newtheorem{prop}[thm]{{\bf Proposition}}

\newtheorem{Ass}[thm]{{\bf Assumption}}

\newtheorem{defn}[thm]{{\bf Definition}}
\newtheorem{rem}[thm]{{\bf Remark}}

\newcounter{Exami}

\newcommand{\Proof}[2][Proof]{
\begin{proof}[{\bf #1}]
#2
\end{proof}
}

\usepackage{amssymb}

\begin{document}

\begin{flushleft}
{ \Large \bf Existence and nonexistence of wave operators for time-decaying harmonic oscillators}
\end{flushleft}

\begin{flushleft}
{\large Atsuhide ISHIDA}\\
{
Department of Liberal Arts, Faculty of Engineering, Tokyo University of Science, 6-3-1 Niijuku, Katsushika-ku,Tokyo 125-8585, Japan\\ 
Email: aishida@rs.tus.ac.jp
}
\end{flushleft}
\begin{flushleft}
{\large Masaki KAWAMOTO}\\
{
Department of Engineering for Production, Graduate School of Science and Engineering, Ehime University, 3 Bunkyo-cho Matsuyama, Ehime, 790-8577. Japan
Japan, \\ 
Email: kawamoto.masaki.zs@ehime-u.ac.jp
}
\end{flushleft}
\begin{abstract}
Controlled time-decaying harmonic potentials decelerate the velocity of charged particles; however, the particles are never trapped by the harmonic potentials. This physical phenomenon changes the threshold between the short-range and long-range classes of the potential through physical wave operators. We herein report that the threshold is $1/(1-\lambda)$ for some $0\leq \lambda<1/2$, as determined using the mass of the particle and a coefficient of the harmonic potential.    

\end{abstract}

\begin{flushleft}

{\em Keywards:} \\ Scattering theory, Nonexistence of wave operators, Harmonic
 oscillator, Time-dependent magnetic fields. \\ 

\end{flushleft}

\section{Introduction}

Let us consider the Hamiltonian 
\begin{align} \label{3}
H_0(t) = \frac{p^2}{2m} + \frac{k(t)}{2} x^2, \quad \mbox{on} \quad L^2({\bf R}^n) , 
\end{align}
where $x = (x_1,x_2,..., x_n) \in {\bf R}^n$, $ p= -i (\partial _1, \partial _2,..., \partial _n)$, and $m>0$ are the position, momentum, and mass of the particle, respectively. The coefficients of harmonic oscillator $k(t)$ belong to $L^{\infty} ({\bf R})$, which converges to $0$ as $t \to \pm \infty$. Define the propagator for $H_0(t)$ by $U_0(t,s)$, that is, a family of unitary operators $\{U_0(t,s)\}_{(t,s) \in {\bf R}^2}$
in $L^2({\bf R}^n)$, whose components satisfy
\begin{align*}
&i \partial _t U_0(t,s) =H_0(t)U_0(t,s) , \quad 
i \partial_s U_0(t,s) = - U_0(t,s) H_0(s), \\ 
&U_0(t,\theta)U_0(\theta ,s) =U_0(t,s), \quad U_0(s,s)= \mathrm{Id}. 
\end{align*} 
Here, owing to the result of Theorem 3 in Fujiwara \cite{F} (see also Yajima \cite{Ya}), the unique existence of $U_0(t,s)$ is evident. 

 Define the observables acting on $C_0^{\infty} ({{\bf R}^n})$ by $$ x_0(t) := U_0(t,0)^{\ast} x
U_0(t,0), \quad p_0(t) := U_0(t,0)^{\ast} p
U_0(t,0).
$$ Then, the commutator calculation shows that
\begin{align*}
 & x' _0(t) = p_0(t) /m, \quad p_0(0) = p, \\ 
 & x''_0(t) + (k(t)/m) x_0(t) = 0, \quad x_0(0) = x , \ x'_0(0) = p/m
\end{align*} 
hold, where $x'_0(t) = (\partial _t x_0)(t) $ and $x''_0(t) = (\partial
_t^2 x_0)(t)$. Define the fundamental solutions $\zeta _1 (t)$ and $\zeta
_2 (t)$ as solutions to 
\begin{align} \label{1}
\zeta _j ''(t) + \left( 
\frac{k(t)}{m}
\right)\zeta _j (t) =0, \quad 
\begin{cases}
\zeta _1 (0) = 1, \\ 
\zeta _1 '(0) = 0,
\end{cases}
\quad 
\begin{cases}
\zeta _2 (0) = 0, \\ 
\zeta _2 '(0) = 1. 
\end{cases}
\end{align}
Therefore, we have $x_0(t) = \zeta _1 (t) x + \zeta _2 (t) p /m$ and $p_0 (t) = m \zeta _1 '(t)x + \zeta _2 '(t) p $ on $C_0^{\infty} ({\bf R}^n)$, see also Kawamoto \cite{Ka}. 

In this study, we assumed the following condition on time-decaying harmonic oscillators:
\begin{Ass} \label{A1} 
Let $r_0 >0$ and $0 \leq \sigma < m/4$, and let the coefficient $k \in L^{\infty}({\bf R}) $ satisfy 
\begin{align} \label{2}
k(t) = \frac{\sigma}{t^2} , \quad \mbox{for all } |t| \geq r_0 > 0.
\end{align}
Moreover, assume that both solutions of
 \eqref{1} with respect to \eqref{2} are
 twice differentiable functions.
\end{Ass} 
An example of $k (t)$ satisfying the assumption
\ref{A1} can be found in \cite{Ka}. A more general case was considered by Geluk--Mari\'{c}-Tomi\'{c} \cite{GMT} (the case of $\lambda =0$ can be found in \cite{Na}). Let us define $\lambda$, $0 \leq \lambda < 1/2 $ as the smaller solution to $\lambda (\lambda -1)
+ \sigma /m = 0 $, i.e.,  
\begin{align*}
\lambda = \frac{1- \sqrt{1-4 \sigma/m}}{2}. 
\end{align*}
Under Assumption \ref{A1}, for $ t > r_0$, it follows that $
 t^{\lambda}$ and $  t^{1- \lambda}$ are linearly independent
solutions to $f''(t) + \CR{\sigma} t^{-2} f(t)/m =0$. Hence, for $t > r_0$, we have for some $c_1,c_2,c_3$ and $c_4 \in {\bf R}$, $\zeta _1(t) = c_1 t^{1- \lambda} + c_2 t^{\lambda}$, $\zeta _2(t) = c_3 t^{1- \lambda} + c_4 t^{\lambda}$. This gives $x_0(t) \phi  = \zeta _1 (t) x \phi  + \zeta _2 (t) p \phi /m = \CAL{O}(t^{1- \lambda})$, and $x_0'(t)\phi  = m\zeta _1 '(t) x \phi + \zeta _2 '(t) p \phi /m=\CAL{O} (t^{- \lambda})$ holds for $ \phi \in C_0^{\infty} ({\bf R}^n)$. Hence, the charged particle is decelerated by the harmonic potential. Nevertheless, we expect that the charged particle will not be trapped owing to the effect by the coefficient $k(t)$. This physical phenomenon was first discovered by \cite{Ka}, and the existence of wave operators and range characterization of wave operators were considered for the case where the potential decays faster than $(1+ |x|)^{-1/(1- \lambda)}$ as $|x| \to \infty$. However, in \cite{Ka}, only the necessary condition for the wave operators to exist was considered and was not considered sufficient. Moreover, the modifier 
and modified wave operators were not discussed. Hence, we herein first introduce the so-called Dollard-type modifier and prove the existence of modified wave operators under Dollard-type long-range potentials. Next, we obtain the sufficient condition of the decaying order of the potential for the wave operators to exist, and observe how the deceleration phenomenon changes the threshold of the decaying order between the short-range and long-range classes of potential $V(t,x)$. We observed that this threshold changed from $1$ to $1/(1- \lambda)$ through the nonexistence of typical wave operators. The potentials are classified into the following three types: 
\begin{defn}\label{A2}
We say that the potential $V^{\mathrm{S}} (t,x)$ belongs to the short-range class if $V^{\mathrm{S}} $ satisfies $V^{\mathrm{S}} \in C({\bf R} ; C ({\bf R}^n)) $ and that for some $\rho_{\mathrm{S}} >1/(1- \lambda)$, there exists $C_{\mathrm{S}} >0$ such that 
\begin{align*}
\left| V^{\mathrm{S}} (t,x) \right| \leq C_{\mathrm{S}} (1+|x|)^{- \rho_{\mathrm{S}}}
\end{align*}  
holds; We say that $V_D^{\mathrm{L}} (t,x)$ belongs to the Dollard-type long-range class if $V_D^{\mathrm{L}}$ satisfies $V_D^{\mathrm{L}} \in C({\bf R}; C^2({\bf R}^n)) $ and that for some $(2 \lambda +1)/ (2(1- \lambda)) < \rho_{D,\mathrm{L}} \leq 1/(1- \lambda) $ and for all multi-index $\alpha$ with $|\alpha| \leq 2$, there exists $C_{D,\mathrm{L},\alpha} >0 $ such that 
\begin{align*}
\left| 
\partial _{x}^{\alpha} V_D^{\mathrm{L}} (t,x)
\right| \leq C_{D, \mathrm{L},\alpha} (1+|{x}|)^{- \rho_{D, \mathrm{L}} -|\alpha|}
\end{align*}
holds; We say that $V^{\mathrm{L}} (t,x)$ belongs to the (typical) long-range class if $V^{\mathrm{L}} (t,x)$ satisfies $V^{\mathrm{L}} \in C({\bf R}; C({\bf R}^n)) $ and that for some $0< \rho_{\mathrm{L}} \leq 1/(1-\lambda) $, there exists $0< C_{\mathrm{L}} \leq \tilde{C}_{\mathrm{L}} $ such that 
\begin{align*}
 C_{\mathrm{L}} |x|^{- \rho_{\mathrm{L}}} \leq V^{\mathrm{L}}(t,x) \leq \tilde{C}_{\mathrm{L}} |x|^{-{\rho}_{\mathrm{L}}} 
\end{align*}
holds for $|x| \geq 1$.
\end{defn}

We now define $V(t)$ as a multiplication operator of $V(t,x) $ and assume that $V(t,x)$ can be written as a linear combination of $V^{\mathrm{S}} (t,x)$, $V^{\mathrm{L}}_D (t,x)$, and $V^{\mathrm{L}} (t,x)$. Subsequently, we define $H(t) := H_0 (t) + V(t)$ and $U(t,s)$ as propagators for $H(t)$. By the result of \cite{F} or \cite{Ya} and the boundedness of $\sup_{t,x} |V(t,x)|$, the unique existence of $U(t,s)$ can be guaranteed.

To prove the existence and nonexistence of wave operators, we decomposed the propagators $U_0(t,0)$ and $U(t,0) $ into simplified propagators. Hence, the following factorization of the propagators was obtained: 
\begin{prop} \label{P41}
Let $A= x \cdot p + p\cdot x$, and for $ \CR{| t|  \geq  r_ 0}$, let $U_{\CR{S,0}} (t, \pm r_ 0)$ and $U_S(t, \pm r_0)$ be propagators for 
\begin{align*}
H_{\CR{S,0}} (t) := \frac{p^2}{2m |t| ^{2\lambda}}, \quad \mbox{and} \quad H_{S} (t) := \frac{p^2}{2m |t|^{ 2\lambda}} +V(t, |t| ^{\lambda} x), 
\end{align*}
respectively. Then, for $\CR{ | t| \geq r_0}$, the following propagator factorizations holds: 
\begin{align*}
U_0(t, \pm r_0) = e^{im \lambda x^2/(2t)}e^{-i \lambda (\log |t|) A/2} U_{\CR{S,0}}(t, \pm r_0)
\end{align*}
and 
\begin{align} \label{20}
U(t, \pm r_0) =e^{im \lambda x^2/(2t)}e^{-i \lambda (\log |t|) A/2} U_{S}(t, \pm r_0). 
\end{align}
\end{prop}
This proposition can be proven by using the approach of Korotyaev \cite{Ko}; we state the proof of this proposition in \S{4}. 

\subsection{Modified wave operators}
By the definition of propagators, for $t> r_0$, we have 
\begin{align*}
U_0(t,0) = U_0(t,r_0) U_0(r_0,0), \quad U(t,0) = U(t,r_0) U(r_0,0), 
\end{align*}
which yields 
\begin{align}\nn
U(t,0)^{\ast} U_0(t,0) &= U(r_0,0)^{\ast} \left( U(t, r_0)^{\ast} 
U_0(t,r_0)
\right) U_0(r_0,0)  
 \\ &= 
 U(r_0,0)^{\ast} \left( U_S(t, r_0)^{\ast} 
U_{S,0}(t,r_0)  \right) U_0(r_0,0). \label{j13}
\end{align}
Hence, in the following, we consider the reduced wave operators 
\begin{align*}
W_S^{\pm} := \mathrm{s-}\lim_{t \to \pm \infty} U_{S}(t,\pm r_0)^{\ast}  U_{S,0} (t, \pm r_0)
\end{align*}
and reduced (Dollard-type) modified wave operators 
\begin{align*}
W_{S,D}^{\pm} := \mathrm{s-}\lim_{t \to \pm \infty} U_{S}(t, \pm r_0)^{\ast}  U_{S,0} (t, \pm r_0)e^{-i \int_{\pm r_0}^t V^{\mathrm{L}}_D (s, s |s| ^{-\lambda}p/\CR{(}m(1-2 \lambda) \CR{)}) ds}
\end{align*}

The main theorems presented herein are as follows: 
\begin{thm} \label{T1}
Assume that $k(t)$ satisfies the assumption \ref{A1}. If $V(t) $ is short range, that is,  $V(t,x) = V^{\mathrm{S}}(t,x) $, then the wave operators $W_{S}^{\pm}$ exist. If $V(t) $ is long range in terms of the Dollard-type modifier, that is, $V(t,x) = V^{\mathrm{S}}(t,x) + V^{\mathrm{L}} _D(t,x)$, then modified wave operators $W_{S,D}^{\pm}$ exist.
\end{thm}
\begin{thm} \label{T2} 
Assume that $k(t)$ satisfies Assumption \ref{A1}. If $V(t,x) =  V^{\mathrm{L}} (t,x)$, then the wave operators $W^{\pm}_{S}$ do not exist. 
\end{thm}
\begin{rem}
Consider the quantum system with time-dependent magnetic fields. The Hamiltonian in such system can be written as  
\begin{align*}
H_B(t) &:= H_{0,B} (t) + V(t ) , \\ 
H_{0,B} (t) &:= 
\frac{1}{2m} \left( 
p_1+ \frac{qB(t)}{2} x_2
\right)^2 + \frac{1}{2m} \left( 
p_2 - \frac{qB(t)}{2} x_1
\right)^2 , \quad \mbox{on} \quad L^2({\bf R}^2), 
\end{align*}
where $q \neq 0$ and $B(t) \in L^{\infty}({\bf R}_t) $ are the charge of the particle and the intensity of the magnetic field, respectively. Let us suppose that $B(t)$ is decaying in $t$. Then, by rewriting 
\begin{align*}
\frac{q^2B(t)^2}{4m} = k(t)
\end{align*}
and by supposing that $k(t)$ satisfies Assumption \ref{A1}, we can obtain almost the same results to those of Theorem \ref{T1} and \ref{T2} for this quantum system, see \cite{Ka}.
\end{rem}

\section{Existence of modified wave operators}

We now prove Theorem \ref{T1}. In this section, we consider only the case of $t>0$. We first prove the following propagation estimation for free propagator $U_{S,0}(t,r_0)$. We set $\chi$ as a characteristic function such that 
\begin{align*}
\chi (s \leq \tau ) := 
\begin{cases}
1 & s \leq \tau, \\ 
0 & s \geq \tau, 
\end{cases}
\quad \chi (s \geq \tau) := 1 - \chi (s \leq \tau).
\end{align*}
\begin{prop}\label{P1} 
Let $\ep>0$ be a sufficiently small constant and $R> \ep $ a large constant. Let $\phi \in \SCR{S}({\bf R}^n)$ and $\hat{\phi} \in C_0^{\infty} ({\bf R}^n \backslash \{ 0\})$ with $\mathrm{supp} (\hat{\phi}) \CR{\subset} \{ \xi \in {\bf R}^n \,|\,  2\ep \leq |\xi| \leq R \} $, where $\hat{\phi}$ denotes the Fourier transform of $\phi$. Then, for $N \in \{ 0,1,2\}$ and for a large $t>r_0$,  
\begin{align} \label{4}
\left\| 
\chi  (|x|/t^{1- 2\lambda} \leq \ep /(m(1-2 \lambda)) ) U_{S,0}(t,r_0) \phi 
\right\| \leq \CR{C_N} t^{-N(1-2\lambda)} \| \J{p^2 + x^2 }^{N/2} \phi  \|
\end{align}
and 
\begin{align} \label{5}
\left\| 
\chi  (|x|/t^{1- 2\lambda} \geq 3R/(m(1-2 \lambda)) ) U_{S,0}(t,r_0) \phi 
\right\| \leq \CR{C_N} t^{\CR{-N (1-2\lambda)}} \| \J{p^2 + x^2 }^{N/2} \phi  \|
\end{align}
hold.
\end{prop}
\Proof{ We only prove \eqref{4} as \eqref{5} can be proven similarly. Let $\ep _0 : = \ep /(m(1-2 \lambda))   $. It is noteworthy that $U_{\CR{S,0}}(t,r_0) = e^{-i t^{1-2 \lambda}p^2/(2m(1-2 \lambda)) } e^{i r_0^{1-2 \lambda} p^2/(2m(1-2 \lambda))}$. Then, by the definition of $\phi$, $\chi(p^2 \leq (3\ep/2)^2) U_{\CR{S,0}}(t,r_0) \phi  \equiv 0$ and $\chi(p^2 \geq (2R) ^2) U_{\CR{S,0}}(t,r_0) \phi  \equiv 0$ hold. Hence, it is sufficient to consider the term $$ \chi  (|x|/t^{1- 2\lambda} \leq \ep_0 )  \chi ((3 \ep/2)^2 \leq p^2 \leq (2R)^2) U_{0,S} (t,r_0) \phi  .$$ For simplicity, $\tilde{\chi} (p^2)$ denotes $ \chi ((3 \ep/2)^2 \leq p^2 \leq (2R)^2)$, and $v(\xi) $ denotes $e^{ir_0^{1-2 \lambda} \xi^2/(2m(1-2 \lambda)) }\hat{\phi} (\xi) $. Then, 
\begin{align*}
&\left\| 
\chi (|x|/t^{1-2 \lambda} \leq \ep _0) \tilde{\chi} (p^2) U_{\CR{S,0}} (t,r_0) \phi
\right\|^2 \\ & = \int_{{\bf R}^n} \left| \int_{{\bf R}^n} \chi (|x|/t^{1-2 \lambda} \leq \ep _0) e^{ix \cdot \xi}e^{-it^{1-2 \lambda}\xi ^2 /(2m(1-2 \lambda)) } \tilde{\chi} (\xi) v(\xi) d \xi \right| ^2 dx
\end{align*}
holds. Because 
\begin{align*}
\left| 
x - \frac{t^{1-2 \lambda} \xi}{m(1-2 \lambda)}
\right| \geq \frac{t^{1-2 \lambda} }{2} \ep_0
\end{align*}
holds on the support of $\chi (|x|/t^{1-2 \lambda} \leq \ep _0)$ and $\tilde{\chi} (\xi)$, we obtain 
\begin{align*}
\CR{\left\| 
\chi (|x|/t^{1-2 \lambda} \leq \ep _0) \tilde{\chi} (p^2) U_{ {S,0}} (t,r_0) \phi
\right\|} \leq C t^{-N(1-2 \lambda)} \| (x^2)^{N/2}e^{ir_0^{1-2 \lambda}p^2/(2m(1-2 \lambda)) }{\phi} \| 
\end{align*} 
through integration by parts. Combined with 
\begin{align*}
\left\| 
(x^{2})^{N/2} e^{ir_0^{1-2 \lambda} p^2/(2m(1-2 \lambda)) }\phi
\right\| \leq C \| (x^{2} + p^{2} )^{N/2} \phi \|, 
\end{align*}
the proposition is thus proven.
} 
\begin{prop}\label{P2} 
Let $\phi \in \SCR{S} ({\bf R} ^n)$ be the same as that in Proposition \ref{P1}. Then, for all $N \in \{0,1,2\}$, 
\begin{align} \label{9}
\left\| 
(-\Delta_{\xi})^{N/2} e^{-i \int_{r_0}^t V^{\mathrm{L}} _D(s, s^{1- \lambda}\xi/(m(1- 2\lambda)) ds} \hat{\phi}  
\right\| \leq \CR{C_N} |{t}|^{N(1-(1-
 \lambda)\rho_{D,\mathrm{L} })}
\end{align}
holds as $t \to \infty$, where we remark that $1-(1-\lambda)\rho_{D,\mathrm{L}}   \CR{\geq} 0$ holds.  
\end{prop}
\Proof{
Let $\alpha (s)$ denotes $s^{1- \lambda}/(m(1- 2\lambda))$. Considering the definition of $V^{\mathrm{L}}_D$, we observed that 
\begin{align*} 
 \left| \CR{(}
\nabla V^{\mathrm{L}}_D \CR{)} (s, \alpha (s) \xi )
\right| \leq 
C \J{s}^{-(1- \lambda)(1 + \rho_{D,\mathrm{L}})}, \quad s \gg 1,
\end{align*}
on the support of $\hat{\phi} (\xi)$. Hence, because 
\begin{align*}
e^{i \int_{r_0}^t V^{\mathrm{L}}_D (s, \alpha (s) \xi ) ds} i\nabla_{\xi}  e^{-i \int_{r_0}^t
 V^{\mathrm{L}} _D(s, \alpha (s) \xi ) ds} = i\nabla _{\xi} + \int_{r_0}^t \alpha  (s) \CR{(} \nabla  V^{\mathrm{L}}  _D \CR{)} (s, \alpha (s) \xi) ds
\end{align*}
holds, we have 
\begin{align}
\left\| 
(-\Delta_{\xi})^{N/2} e^{-i \int_{r_0}^t V^{\mathrm{L}} _D(s, \alpha (s) \xi) ds} \hat{\phi}  
\right\| = \left\| 
\left( 
i \nabla _{\xi} + \int_{r_0}^t \alpha (s) \CR{(} \nabla  V^{\mathrm{L}} _D \CR{)} (s, \alpha (s) \xi) ds
\right) ^{N} \hat{\phi}
\right\|. \label{8}
\end{align}
Let $k$ and $l$ be multi-indexes with $|k|, |l| \leq N$. Then, it follows that $|\alpha (s) | ^{|k|} |(\partial_{\CR{x}}^k V^{\mathrm{L}} _D)(s, \alpha (s) \xi)| = \CAL{O} (s^{-(1- \lambda) \rho_{D,\mathrm{L}}})$
holds on the support of $(\CR{\partial} _{\xi}^{l} \hat{\phi})(\xi) $. Hence, we obtain  
\begin{align*}
\eqref{8} \leq C_N \left(\int_{r_0}^t s^{-(1- \lambda)\rho_{D,\mathrm{L}}} ds \right) ^{N} \times  \sum_{j=0}^{N} \sum_{|\alpha| \leq j }\left\| \partial _{\xi}
 ^{\alpha } \hat{\phi} \right\|.
\end{align*}
 The inequality above yields \eqref{9}.

}

\begin{prop}\label{P3}
Let $\phi \in \SCR{S} ({\bf R} ^n)$ and $\ep >0$ be the same as those in Proposition \ref{P1}. Then,  
\begin{align} \nn  
& \left\| \chi (|x|/t^{1-2 \lambda} \leq \ep /(m(1-2\lambda))) U_{S,0} (t,r_0) e^{-i \int _{r_0}^t
 V^{\mathrm{L}}_D (s, s^{1- \lambda} \CR{p} /(m(1-2 \lambda)) ds} \CR{\phi} \right\|
 \\ &\leq C |{t}|^{ \CR{2(2 \lambda - (1- \lambda) \rho_{D,L}) } }  \label{10}
\end{align}
holds, as $t \to \pm \infty$.
\end{prop}
\Proof{ 
By Proposition \ref{P2}, 
\begin{align*}
& \left\| \J{p^2 + x^2}  e^{-i \int _{r_0}^t
 V^{\mathrm{L}}_D (s, s^{1- \lambda}p/(m(1-2 \lambda)) ds} \phi \right\|  \\ & 
 =  \left\| \J{ \xi^2 + (i \nabla _{\xi})^2}  e^{-i \int _{r_0}^t
 V^{\mathrm{L}}_D (s, s^{1- \lambda}\xi/(m(1-2 \lambda)) ds} \hat{\phi}(\xi) \right\|_{L^2({\bf R}^n_{\xi})} 
 \leq C |{t}| ^{2(1-(1-
 \lambda)\rho_{\CR{D, \mathrm{L}}}}
\end{align*}
holds. Equation \eqref{10} follows from this estimation and Proposition \ref{P1}.
}
\begin{rem}
By the definition of $\rho_{D,\mathrm{L}}$, it holds that $2 \lambda - (1- \lambda)\rho_{D, \mathrm{L}} < 2 \lambda  - \lambda -1/2 <0$.
\end{rem}

\subsection{Proof of theorem \ref{T1}}
Here, we will prove Theorem \ref{T1}. The function space $\Omega_{0} := \{ \phi \in \SCR{S} ({\bf R}^n)\, | \, \SCR{F}[\phi] \in C_0^{\infty} ({\bf R}^n \backslash \{ 0\}) \} $ is dense on $L^2({\bf R}^n)$, where $\SCR{F}$ indicates the Fourier transform. Therefore, to prove the
existence of modified wave operators, it is sufficient to prove
the existence of 
\begin{align*}
\lim_{t \to \infty} U_{S}(t,r_0)^{\ast}  U_{0,S} (t,r_0)e^{-i\alpha (t,p)} \phi 
\end{align*} 
for all $\phi \in \Omega _0$\CR{, where \begin{align*} 
   \int _{r_0}^t
 V^{\mathrm{L}} _D (s, s^{1- \lambda} p/(m(1- 2\lambda))) ds &= \alpha (t,p)  . 
\end{align*}}
Here, we observed that for all $\phi \in \Omega _0$, there exists $\ep = \ep (\phi) >0$ and large constant $R = R(\phi) > \ep$ such that 
\begin{align*}
\mathrm{supp} \{ \SCR{F}[\phi] (\xi) \} \CR{\subset} \{ \xi \in {\bf R}^n \, | \, 2\ep \leq |\xi| \leq R \}
\end{align*}
holds. For such an $\ep$ and for $0< \delta_0 \ll \ep$, we define $F_{\delta _0} \in C^{\infty} ({\bf R})$ as follows: 
\begin{align*}
F_{\delta_0}( s \leq \tau) :=
\begin{cases}
1 & s \leq \tau - \delta_0, \\ 
0 & s >  \tau, 
\end{cases} 
\quad 
F_{\delta_0} (s \geq \tau) := 
\begin{cases} 
1 & s \geq \tau + \delta_0, \\ 
0 & s < \tau. 
\end{cases}
\end{align*}
Such functions are used as smooth cut-off functions. For simplicity, we use the notations 
\begin{align*}
 F_{\delta _0} (|x| /t^{1-2 \lambda} \leq \ep/(m(1-2 \lambda))) &= F_1(|x|/t^{1-2 \lambda}), \\ 
   1-F_1  &= F_2(|x|/t^{1-2 \lambda}) \CR{.} \end{align*}
where we remark that the support of $F_2(|x|/t^{1-2 \lambda} )$ is equivalent to the support of $F_{\delta_0} (|x| /t^{1-2 \lambda} \geq \ep/(m(1-2 \lambda)) -\delta_0)$. Hereinafter, we assume $t >0$ is sufficiently large. \CR{By $F_1 (|x| /t^{1-2\lambda}) = F_1 (|x| /t^{1-2\lambda}) \chi (|x|/t^{1-2\lambda} \leq \ep/m(1-2 \lambda)  )$ and Proposition 4}, we can obtain 
\begin{align*}
\lim_{t \to \infty} U_S(t,r_0)^{\ast} F_1(|x|/t^{1-2 \lambda})
U_{S,0} (t,r_0) e^{-i \alpha (t,p)} \phi = 0. 
\end{align*} 
Here, we define $\CAL{K} (t) :=  U_S(t,r_0)^{\ast} \CR{F_2(|x|/t^{1-2 \lambda}) }
U_{S,0} (t,r_0) e^{-i \alpha (t,p)} \phi $ and
obtain, for $t_1, t_2 >0$ and $\phi \in \Omega _{0}$,  
\begin{align*}
\CAL{K}(t_1) \phi - \CAL{K} (t_2) \phi  = \int_{t_1}^{t_2} \CAL{K} '(t) \phi  dt. 
\end{align*}
Hence, if one obtains 
\begin{align}\label{13}
 \left\| \CAL{K} ' (t)  \phi \right\| \in L^1((\tilde{R}, \infty ) ; dt )
 \end{align}
for a sufficiently large $\tilde{R} >0$, one can prove that $\CAL{K}(t) \phi $ is a Cauchy sequence by the $L^2
$-norm sense and that $ \lim_{t \to \infty} \CAL{K} (t)$ exists. This approach is called the Cook--Kuroda method. Here, we define the Heisenberg derivative of $F_2$ associated with $H_{0,S} (t)$ by
\begin{align*}
{\bf D}_{{H}_{\CR{S,0}} (t)} \left( F_2\right) := \frac{\partial}{\partial t} F_2(|x|/t^{1-2 \lambda}) + i[H_{\CR{S,0}}(t), F_2(|x|/t^{1-2 \lambda})] .
\end{align*}
Straightforward calculations show 
\begin{align*}
{\bf D}_{{H}_{\CR{S,0}} (t)} \left( F_2 \right) &= \frac{-i}{\CR{2}mt^{2(1-2 \lambda) + 2 \lambda}} F_2''(|x|/t^{1-2 \lambda}) +
 \frac{xF_2'(|x|/t^{1-2 \lambda})}{mt |x|} \cdot \left( 
p- \frac{m(1-2 \lambda)x}{t^{1-2 \lambda}}
\right) \\ & \quad \CR{ -i \frac{n-1}{2m t |x|} F_2'(|x|/t^{1-2 \lambda}) } ,  
\end{align*}
where we denote $F_2'(|x|/t^{1-2 \lambda})$ and $F_2''(|x|/t^{1-2 \lambda})$ by 
\begin{align*}
\left( \frac{d}{ds}F_{\delta_0}(s \leq \ep/(m(1-2\lambda)))
\right) \Big|_{s=|x|/t^{1-2\lambda}}, \quad \left( \frac{d^2}{ds^2}F_{\delta_0}(s \leq \ep/(m(1-2\lambda)))
\right) \Big|_{s=|x|/t^{1-2\lambda}}.
\end{align*}
Furthermore, we have 
\begin{align*}
& \left\| 
 \frac{xF_2'(|x|/t^{1-2 \lambda})}{mt |x|} \cdot \left( 
p- \frac{m(1-2 \lambda)x}{t^{1-2\lambda}}
\right) U_{\CR{S,0}} (t,r_0) e^{-i \alpha (t,p)} \phi
\right\| \\ & \leq C
t^{-2+ 2 \lambda} \sum_{j=1}^{n} \left\| F_2'(|x|/t^{1-2 \lambda}) 
U_{\CR{S,0}} (t,r_0) \CR{   
\left( x_j - \frac{r_0^{1-2 \lambda} p_j}{m(1-2 \lambda)} 
\right)   e^{- i\alpha (t,{ p})}   {\phi}  } \right\|. 
\end{align*}
Because 
$\mathrm{supp}(F_2') \subset \left\{ \CR{ \ep/(m(1-2 \lambda) - \delta _0 \leq } |x|/t^{1-2 \lambda} \right\}$ hold\CR{s}; therefore, we obtain  
\begin{align}\label{11}
& \left\| {\bf D}_{{H}_{\CR{S,0}} (t)} (F_2) U_{\CR{S,0}} (t,r_0)e^{-i\alpha (t,p)} \phi  \right\|  
\\ \nn & \leq 
Ct^{-2(1-2 \lambda)-2 \lambda } +   C t^{-2 + 2 \lambda }t^{1-(1- \lambda)\rho_{D, \mathrm{L}}  } 
\in L^1(\CR{(\tilde{R}, \infty ) ;dt})
\end{align}
using 
Proposition \ref{P2}. Moreover, it is easy to obtain 
\begin{align}\label{12}
 \left\| V^{\mathrm{S}} (t, t^{\lambda}x)) F_2(\CR{|x|/t^{1-2 \lambda}}) U_{\CR{S,0}}(t,r_0)e^{-i\alpha (t,p)} \phi  \right\|  \in L^1( \CR{ (\tilde{R} , \infty) ; dt })
\end{align}
for all $\psi \in L^2({\bf R}^n)$ because $\left\|  V^{\mathrm{S}}
(t,t^{\lambda} x) F_2(|x|/t^{1-2 \lambda}) \psi \right\| \leq C |t| ^{-(1-\lambda)
\rho_{\mathrm{S}}}$ holds under the definition of $V^{\mathrm{S}}$. Hence, the proof becomes complete if we have 
\begin{align*}
\left\| 
F_2(|x|/t^{1-2\lambda} ) \left( 
V_D^{\mathrm{L}} (t,t^{\lambda}x ) - V_D^{\mathrm{L}} (t,t^{1-\lambda} p/m(1-2 \lambda))
\right)  U_{\CR{S,0}}(t,r_0)e^{-i\alpha (t,p)} \phi
\right\| \in L^1(\CR{(\tilde{R},\infty)};dt), 
\end{align*}
which will be proven in the following.

We define 
\begin{align*}
&\tilde{F}_2(t,x) = F_{\delta_0} \left( 
|x|/ t^{1-2 \lambda} \geq \ep/(m(1-2\lambda)) - 2\delta_0
\right), \quad 
\tilde{\varphi} (\xi ^2)= F_{\delta_0}  \left( 
|\xi|^2 \geq  4\ep ^2-  \delta_0
\right), \\ 
&F_3(t;s) = F_{\delta _0} \left( 
|s|/t^{1- \lambda} \geq \delta 
\right), \quad 0< {\delta + 2\delta_0}  < \frac{\mathrm{min} \left\{ 
\ep , \sqrt{4 \ep^2 -\delta_0}
\right\}}{({m(1-2 \lambda)})}. 
\end{align*}
They satisfy the following conditions: 
\begin{align} \nn
& F_2 \tilde{F}_2 = F_2,  \quad \tilde{\varphi} (\xi ^2) \hat{\phi} (\xi) =
 \hat{\phi} (\xi), \\ 
 & \tilde{F}_2(t,x) F_3 (t; t^{\lambda} x
) = \tilde{F}_2(t,x), \quad \mbox{and} \quad \tilde{\varphi} (p^2) F_3 (t;t^{1- \lambda}p/m(1-2 \lambda)) =
\tilde{\varphi} (p^2). \label{Ad1}
\end{align}
 By \eqref{11}, \eqref{12}, and \eqref{Ad1}, we found that $\CR{\CAL{K}'(t) } \phi $ can be calculated by 
\begin{align*}
\CAL{K}'(t) \phi  &=  U_S(t,r_0)^{\ast} i F_2(|x|/t^{1-2 \lambda}) \\ & \quad \times \left( 
 V^{\mathrm{L}}_D (t,t^{\lambda} x)\tilde{F}_2(|x|/t^{1-2 \lambda}) - V^{\mathrm{L}}_D (t, t^{1- \lambda} p/(m(1- 2\lambda))) \tilde{\varphi } (p^2) 
\right) \\ & \qquad \times   U_{0,S} (t,r_0)e^{-i
 \alpha (t,p)} \phi +\CAL{O}(t^{-(1-\lambda) \rho _{\mathrm{S}}}) + \CR{\CAL{O}(t^{-2+ 2 \lambda}) + \CAL{O} (t^{-1 + 2 \lambda - (1- \lambda) \rho_{D, \mathrm{L}} } )} .
\end{align*}
Next, we express
\begin{align*}
V_{\mathrm{L}} (t; \cdot ) := V^{\mathrm{L}}_D (\cdot ) F_3(t; \cdot) 
\end{align*}
and found that 
\begin{align} \label{Ad3}
\sup_{y \in {\bf R}^n }\left| 
(\nabla^{\alpha} _y V_{\mathrm{L}})(t; y )
\right| \leq C |{t}|^ {-(|\alpha | + \rho_{D,\mathrm{L}}) (1-
 \lambda)} 
\end{align}
holds. As the term $  
 V^{\mathrm{L}}_D (t,t^{\lambda}x)\tilde{F}_2 - V^{\mathrm{L}} _D(s ,s^{1-\lambda}p/(m(1-2 \lambda))) \tilde{\varphi } (p^2) 
$ can be rewritten by $ \CR{(} V_{\mathrm{L}} (t;t^{\lambda}x) - V_{\mathrm{L}} (t; t^{1- \lambda}p/(m(1-2 \lambda))) \CR{) \tilde{\varphi} (p^2) }$, we prove that
\begin{align} \label{14} \left\| 
\left( 
V_{\mathrm{L}} (t; t^{\lambda}x) - V_{\mathrm{L}} (t,t^{1- \lambda}p/(m(1-2 \lambda)))
\right) U_{S,0}(t,r_0) e^{-i \alpha (t,p)} \phi \right\| \in L^1(\CR{(\tilde{R}, \infty) ;dt })
\end{align}
holds in the following. Then, the existence of modified wave operators can
be proven using \eqref{11}, \eqref{12}, and \eqref{14}. Using the Baker--Campbell--Hausdorff formula, we have 
\begin{align}
& \nn   V_{\mathrm{L}} (t; t^{\lambda} x) - V_{\mathrm{L}} (t; t^{1- \lambda}p/(m(1-2 \lambda)) ) 
\\ &= \label{15}-
\frac{ti }{m(1-2 \lambda)} \int_0^1 (1-\tau)(\Delta V_{\mathrm{L}})
 (t; \tau(t:x,p) ) d \tau \\
 &  + t^{\lambda}
\int_0^1 (\nabla V_{\mathrm{L}}) (t;\tau(t;x,p) ) d \tau \cdot \left( 
x- \frac{t^{1-2 \lambda} }{m(1-2 \lambda)}p
\right) \label{16}, \\ 
\nn &\tau  (t;x,p) :=  \tau (t^{\lambda}x - t^{1- \lambda}p/(m(1-2 \lambda))) + t^{1- \lambda}
 p/(m(1-2 \lambda)) . 
\end{align}
By \eqref{Ad3}, we have
\begin{align}\label{Ad4}
t \left\| 
\int_0^1 |(\Delta V_{\mathrm{L}})
 (t;  \tau(t:x,p) )| d \tau 
\right\| \leq C t^{1-(2+  \rho_{D,\mathrm{L}} )(1- \lambda)}.
\end{align}
Because $1-(2+  \rho_{D,\mathrm{L}} )(1- \lambda) < -1 + 2 \lambda -(2 \lambda +1)/2 = -3/2 + \lambda < -1 $, \eqref{Ad4} is integrable in $t$. Next, we \CR{estimate \eqref{16} and prove
\begin{align} \label{Ap}
\left\| 
 t^{\lambda}
\int_0^1 (\nabla V_{\mathrm{L}}) (t;\tau(t;x,p) ) d \tau \cdot \left( 
x- \frac{t^{1-2 \lambda} }{m(1-2 \lambda)}p
\right) U_{S,0} (t,r_0) e^{-i \alpha (t,p)} \phi 
\right\| 
\end{align}
} is integrable in $t$. By the definition of
$V_{\mathrm{L}} ( t;\cdot )$, 
\begin{align*} 
\left\| 
(\partial_j V_{\mathrm{L}})(t; \CR{\tau (t; x,p)} )
\right\| \leq C \CR{t}^{- \CR{(1+ \rho_{D, \mathrm{L}} )} (1-
 \lambda)}
\end{align*}
holds; additionally, by \eqref{9},  
\begin{align*}
\left\|  
\left( 
x_j - \frac{t^{1-2 \lambda} }{m (1-2 \lambda)} p_j
\right)  U_{0,S} (t,r_0) e^{-i \alpha (t,p)} \phi
\right\| \leq C \CR{{t} ^{1- (1- \lambda) \rho_{D,\mathrm{L}}}},
\end{align*}
holds for all $j \in \{1,...,n\}$. Hence, we obtain 
\begin{align*}
& t^{\lambda}
\left\| \int_0^1 (\nabla V_{\mathrm{L}}) (t;\tau(t;x,p) ) d \tau \cdot \left( 
x- \frac{t^{1-2 \lambda} }{m(1-2 \lambda)}p
\right) U_{0,S} (t,r_0) e^{-i \alpha (t,p)} \phi
\right\|  \\ & 
\leq C \CR{t}^{ 2 \lambda - 
 2 (1- \lambda) \rho_{D,\mathrm{L}}}. 
\end{align*}
Considering the definition of $\rho_{D,\mathrm{L}}$, $2 \lambda -2(1- \lambda) \rho_{D,\mathrm{L}} < -1$ holds. Combined with \eqref{Ad4}, we have \CR{$\eqref{Ap} \in L^1((\tilde{R}, \infty) ;dt)$}. This proves \eqref{13}.

\section{Nonexistence of wave operators}
In this section, we prove Theorem \ref{T2}. The nonexistence of wave operators for long-range perturbations was first considered by Dollard \cite{Do}, and this approach was applied to some Hamiltonians, see e.g., Ozawa \cite{O}, Jensen-Ozawa \cite{JO}, Ishida \cite{I}, \cite{I2}. In particular, we refer to the approach of Ozawa \cite{O}. Let $(\cdot, \cdot)$ denote the inner products on $L^2({\bf R}^n)$ and $t \geq s > r_0 \CR{>0}$. In the proof, we assume that that transformed wave operator 
\begin{align*}
\tilde{W}^+ = \mathrm{s-} \lim_{t \to \infty} U_S(t,r_0)^{\ast} e^{-it^{1-2 \lambda}p^2/\CR{(2m(1-2 \lambda)) }} e^{-im(1-2 \lambda) x^2/(2t^{1-2 \lambda})}  
\end{align*}
exists for $V= V^{\mathrm{L}}$ and causes a contradiction. If $\tilde{W}^+$ does not exist, then the typical wave operator
\begin{align*}
 \CR{W_S^{+}}  &= \mathrm{s-} \lim_{t \to \infty} U_S(t,r_0)^{\ast} U_{S,0}(t,r_0) \\ 
&= \mathrm{s-} \lim_{t \to \infty} U_S(t,r_0)^{\ast} e^{-it^{1-2 \lambda}p^2/\CR{(2m(1-2 \lambda))} } e^{-im(1-2 \lambda) x^2/(2t^{1-2 \lambda})} \\ & \qquad \qquad \times e^{im(1-2 \lambda) x^2/(2t^{1-2 \lambda})} e^{ir_0^{1-2 \lambda} p^2/ \CR{(2m(1-2 \lambda))}}
\end{align*}
does not exist as well because the operator $e^{im(1-2 \lambda) x^2/\CR{(2t^{1-2 \lambda})}} e^{ir_0^{1-2 \lambda} p^2/\CR{(2m(1-2 \lambda))}}$ is uniformly bounded in $t$. To consider the nonexistence of $\tilde{W}^+$, we introduce the following modified propagation estimation: 
\begin{prop}\label{P4}
Suppose that $\phi \in C_0^{\infty} ({\bf R}^n \backslash \{ 0 \})$ with $\mathrm{supp}(\hat{\phi}) \subset \{\xi \in {\bf R}^n \, | \, |\xi| \geq 2m(1-2 \lambda) \ep \}  $ for some $\ep >0$. Then, 
\begin{align} \label{Ad5}
\left\| 
\chi (|x|/t^{1- 2\lambda} \leq  \ep ) e^{-it^{1-2 \lambda}p^2/\CR{(2m(1-2 \lambda) )}} e^{-im(1-2 \lambda) x^2/(2t^{1-2 \lambda})} \phi 
\right\| = 0
\end{align} 
holds.
\end{prop}
\Proof{
For simplicity, we define 
\begin{align*}
\Lambda := m(1-2 \lambda).
\end{align*}
Furthermore, we define the observable 
\begin{align*}
\tilde{x}(t) :=  e^{i\Lambda x^2/(2t^{1-2 \lambda})}e^{it^{1-2 \lambda}p^2/ \CR{(2\Lambda)} } x e^{-it^{1-2 \lambda}p^2/ \CR{(2\Lambda )} } e^{-i\Lambda x^2/(2t^{1-2 \lambda})}. 
\end{align*}
By simple calculations, it follows that for all $a$, $b \in {\bf R} \backslash \{ 0\}$, 
\begin{align*}
e^{ia x^2} \MAT{x\\p} e^{-ia x^2} = \MAT{x \\ p-2ax}, \quad 
e^{ib p^2} \MAT{x\\p} e^{-ib p^2} = \MAT{x+2bp \\ p}. 
\end{align*}
These equations yield 
\begin{align}\label{Ad6}
\tilde{x} (t) = e^{i\Lambda x^2/(2t^{1-2 \lambda})} \left( x + t^{1-2 \lambda}p/\Lambda \right) e^{-i\Lambda x^2/(2t^{1-2 \lambda})} = t^{1-2 \lambda}p/\Lambda. 
\end{align}
Hence, the left hand side of \eqref{Ad5} is equal to 
\begin{align*}
\left\| 
\chi (|t^{1-2 \lambda} \xi |/\Lambda t^{1- 2\lambda} \leq  \ep )  \hat{\phi} (\xi) 
\right\|_{L^2({\bf R}^n_{\xi})} = \left\| \chi (|\xi|/\Lambda \leq \ep) \hat{\phi}(\xi) \right\|_{L^2({\bf R}^n_{\xi})}. 
\end{align*}
Clearly, the support of $\chi (|\xi|/\Lambda \leq \ep)$ and $\hat{\phi}(\xi) $ are disjoint and this proves \eqref{Ad5}. 
 
}

\subsection{Proof of theorem \ref{T2}}
Let $\phi$ satisfies the conditions of Proposition \ref{P4}. Moreover, let us define
\begin{align*}
{U}_0(t) := e^{-it^{1-2 \lambda}p^2/2\Lambda } e^{-i\Lambda x^2/(2t^{1-2 \lambda})} \CR{.} 
\end{align*}
Then, 
\begin{align*}
& \left( 
\left( 
U_{S} (t, r_0) ^{\ast} U_{0} (t) - U_{S} (s,r_0)^{\ast} U_{0} (s)
\right) \phi , \tilde{W}^{+} \phi 
\right)  \\ & = 
\int_s^t \partial _{\tau} \left( 
U_{S}(\tau , r_0) ^{\ast} U_{0} (\tau ) \phi, \tilde{W} ^{+} \phi
\right)  d \tau \\ & = 
i\int_s^t  \left( 
U_{S}(\tau , r_0) ^{\ast} V(\tau, \tau^{\lambda}x) U_{0} (\tau ) \phi, \tilde{W} ^{+} \phi
\right)  d \tau  \\ & \quad  + 
\frac{i\Lambda (1-2 \lambda) }{2} \int_s^t \left( 
U_S(\tau,r_0) U_0(\tau) (x^2 \tau ^{-2+2\lambda}) \phi,  \tilde{W} ^{+} \phi
\right) d \tau
 \\ 
& \equiv I_1 +I_2 +I_3
\end{align*}
with 
\begin{align*}
I_1 &:= i \int_s^t \left( 
U_{0}(t)^{\ast} V(\tau,\tau^{\lambda} x) U_{0} (\tau)\phi, \phi 
\right)  d \tau \\ 
I_2 &:= i \int_s^t 
\left( 
V(\tau, \tau ^{\lambda} x) U_{0} (\tau) \phi , 
\left[ U_S(\tau, r_0) \tilde{W}^{+} 
- U_{0} (\tau) 
\right] \phi 
\right)  d \tau
\end{align*}
and 
\begin{align*}
I_3 := \frac{i\Lambda (1-2 \lambda)}{2} \int_s^t \left( 
U_S(\tau,r_0) U_0(\tau) (x^2 \tau ^{-2+2\lambda}) \phi,  \tilde{W} ^{+} \phi
\right) d \tau.
\end{align*}
 First, we estimate $I_1$: Additionally, we assume $\mathrm{supp} \{ \hat{\phi} \} \subset \{\xi \in {\bf R}^n \, | \, |\xi| \leq \Lambda R \}$ for some $R>0$. Then, by considering \eqref{Ad6} and the definition of $V(\tau, \tau^{\lambda} x)$, for a large $ s$ such that $2s^{1-\lambda}\ep \geq 1 $ and $t \geq s$, it holds that 
 \begin{align} \nn
|  I_1 | &=  \left| \int_s^t \left( 
 U_{0} (\tau) ^{\ast} V(\tau, \tau ^{\lambda} x)  U_{0} (\tau) \phi, \phi
\right)  d \tau \right|  = 
\left| 
\int_s^t \int_{{\bf R}^n} V(\tau, \xi \tau ^{1-\lambda} /\Lambda ) \left| \hat{\phi} (\xi) \right|^2 d \xi d \tau
\right| \\ &= \nn
\left| 
\int_s^t \int_{2\Lambda \ep \leq |\xi| \leq \Lambda R} V(\tau, \xi \tau^{1-\lambda} /\Lambda ) \left| \hat{\phi} (\xi) \right|^2 d \xi d \tau
\right| \\ & \geq 
C_{\mathrm{L}}R^{-\rho_{\mathrm{L}}} \left\| \phi  \right\| ^2 \int_s^t \tau^{-(1-\lambda)\rho_{\mathrm{L}}} d \tau  \label{17}
\end{align}
Next, we estimate $I_2$. By Proposition \ref{P4}, one has 
 \begin{align*}
 & \left\| 
 V(\tau, \tau ^{\lambda}x) U_{0} (\tau) \phi 
 \right\| \\ & \leq \left\| 
 V(\tau, \tau ^{\lambda}x) (1-\chi (|x|/\tau^{1-2 \lambda}\leq \ep )) U_{0} (\tau) \phi 
 \right\| +  \left\| 
 V(\tau, \tau ^{\lambda}x) \chi (|x|/\tau^{1-2 \lambda}\leq \ep )U_{0} (\tau) \phi 
 \right\|  \\ & \leq 
 \tilde{C}_{\mathrm{L}} \ep  ^{-\rho_{\mathrm{L}}}\tau^{-(1- \lambda) \rho_{\mathrm{L}}} \| \phi \|  + 0
 \end{align*}
 Because we assume that $\tilde{W}^+$ exists, retaking a sufficiently large $s$ (if necessary), we can obtain 
 \begin{align*}
 \left\| 
\left(  U_S(\tau,r_0)^{\ast} U_{0}(\tau) - \tilde{W}^{+} \right) \phi 
 \right\| \leq \delta \|\phi \|
 \end{align*}
 for a sufficiently small $0< \delta \ll 1 $. Hence, we select $2\delta = (\ep/R)^{\rho_{\mathrm{L}}} (C_{\mathrm{L}}/\tilde{C}_{\mathrm{L}})$; for some $C>0$, we obtain 
 \begin{align} \label{18}
 |I_2| \leq \frac{C_{\mathrm{L}} \CR{ \| \phi \| ^2}}{2 R^{\rho_{\mathrm{L}}}} \int_s^t \tau ^{-(1- \lambda) \rho_{\mathrm{L}}} d \tau. 
 \end{align} 
 Finally, we estimate $I_3$ as follows: 
 \begin{align}\nn
 |I_3| &\leq \frac{\Lambda (1-2 \lambda) }{2} \left\| x^2 \phi \right\| \left\| \tilde{W}^+ \phi  \right\| \int_s^t \tau ^{-2 + 2 \lambda} d \tau \\ 
 & \leq \frac{\Lambda (1-2 \lambda)}{2} \left\| x^2 \phi \right\| \left\| \tilde{W}^+ \phi  \right\| \int_s^{\infty} \tau ^{-2 + 2 \lambda} d \tau = \frac{ \Lambda (1-2 \lambda) s^{-1+2 \lambda}}{2} \nn\left\| x^2 \phi \right\| \left\| \tilde{W}^+ \phi  \right\| \\ 
 & \leq \left\| x^2 \phi \right\| \left\| \tilde{W}^+ \phi  \right\|, \label{Ad7}
 \end{align} 
 where we use a sufficiently large $s$ (if necessary) such that $\Lambda (1-2 \lambda) s^{ -1+ 2\lambda}/2 \leq 1$. 
 
 Inequalities \eqref{17}, \eqref{18}, and \eqref{Ad7} provide 
 \begin{align*}
 \CR{2}\left\| (1+x^2) \phi \right\| \left\| \tilde{W}^{+} \phi \right\| &\geq \CR{2} \left\| \phi \right\| \left\| \tilde{W}^{+} \phi \right\| \geq |I_1| - |I_2| -|I_3| \\ & \geq 
 \frac{C_{\mathrm{L}} \CR{ \| \phi \| ^2} }{2 R^{\rho_{\mathrm{L}}}} \int_s^t \tau ^{-(1- \lambda) \rho_{\mathrm{L}}} d \tau  - \left\| x^2 \phi \right\| \left\| \tilde{W}^+ \phi  \right\|.
 \end{align*}
This inequality yields  
\begin{align*}
C & \geq  \left\| (1+x^2) \phi \right\| \left\| \tilde{W}^{+} \phi \right\|  \geq 
\frac{C_{\mathrm{L}} \CR{ \| \phi \| ^2}}{4 R^{\rho_{\mathrm{L}}}} \int_s^t \tau ^{-(1- \lambda) \rho_{\mathrm{L}}}  d \tau \to \infty
\end{align*} 
as $t \to \infty$ because $(1- \lambda) \rho_{\mathrm{L}} \leq 1$. Consequently, we have the contradiction that $\tilde{W}^+$ exists, which proves Theorem \ref{T2}. 
 
 \section{Fundamental results}
 
 In this section, we provide some proofs for the fundamental results, which appeared in the reduction process of the propagator.
 
\subsection{Proof of Proposition \ref{P41}}
In this subsection, we prove Proposition \ref{P41}. First, we prove that for \CR{$j \in \{ 1,2,...,n \}$ and} $\beta \in {\bf R}$ with $\beta \neq 0$, 
\begin{align} \label{19}
e^{-i\beta A} \CR{x_j} e^{i \beta A} = e^{\CR{-2 \beta} } \CR{x_j}, \quad  e^{-i\beta A} \CR{p_j} e^{i \beta A} = e^{\CR{2 \beta }} \CR{p_j}
\end{align}
holds. We consider only the term associated to $x_j$. By simple calculations, it follows that
\begin{align*}
\frac{d}{d \CR{\beta}} e^{\CR{-i\beta A}} \CR{x_j} e^{\CR{i \beta A}} = \CR{-} e^{\CR{-i \beta A}} i [A, \CR{x_j}] e^{\CR{i  \beta A}}= \CR{- 2} e^{\CR{-i \beta A}} \CR{x_j} e^{\CR{i  \beta A}}, 
\end{align*}
where $[A_0,B_0]$ denotes the commutator of self-adjoint operators $A_0$ and $B_0$. The equation above with initial condition $e^{\CR{-i  A \times 0}} \CR{x_j} e^{\CR{i  A\times  0}} = \CR{x_j}$ implies that the first one of \eqref{19} holds. \CR{In particular 
\begin{align*}
e^{-i \beta A} x^2 e^{i \beta A} = \sum_{j=1}^n e^{-i\beta A} x_j e^{i \beta A} e^{-i\beta A} x_j e^{i \beta A}  = e^{-4 \beta } x^2
\end{align*}
and 
\begin{align*}
e^{-i \beta A} p^2 e^{i \beta A} = \sum_{j=1}^n e^{-i\beta A} p_j e^{i \beta A} e^{-i\beta A} p_j e^{i \beta A}  = e^{4 \beta } p^2
\end{align*}
hold.} By \eqref{19},  
\begin{align*}
e^{-i \lambda \log |t| A/2} H_{S}(t) &= \left( e^{-i \lambda \log |t| A/2} H_S(t) e^{i \lambda \log |t| A/2} \right)e^{-i \lambda \log |t| A/2} \\ &= \left(  \frac{p^2}{2m} +V(t,x)  \right) e^{-i \lambda \log |t| A/2} .
\end{align*}

Next, we prove \eqref{20}. We denote $\tilde{U} (t,r_0)$ as the right-hand side of \eqref{20}. Then, for all $t \geq r_0$
\begin{align*}
i \frac{d}{dt} \CR{\tilde{U}} (t,r_0) &= \left(  \frac{m \lambda}{2t^2} x^2 + e^{im \lambda x^2/(2t)} \left( 
\frac{\lambda}{2t} A \right) e^{-m \lambda x^2/(2t)}
\right)  \CR{ \tilde{U}} (t,r_0) \\ & \qquad  + 
 e^{m \lambda x^2/(2t)} \left( 
\frac{1}{2m } p ^2 + V(t,  x)
\right)e^{-m \lambda x^2/(2t)} \CR{\tilde{U}}(t,r_0)
\\
& = \Bigg[ 
 \frac{m \lambda}{2t^2} x^2 + \frac{\lambda}{2t}\left(x \cdot \left(p -\frac{m\lambda x}{t}\right) + \left(p -\frac{m\lambda x }{t}\right) \cdot x \right) \\ & \qquad + \frac{1}{2m} \left( 
 p -\frac{m\lambda x}{t}
 \right) ^2 +V(t,x)
\Bigg] \CR{ \tilde{U} }(t,r_0) \\ 
&= \left( 
\frac{1}{2m} p^2 + \frac{k(t)}{2} x^2 + V(t,x)
\right) \CR{\tilde{U}}(t,r_0) = H(t)  \CR{\tilde{U}}(t,r_0)
\end{align*}
holds, where we use $m(\lambda-\lambda ^2) = \CR{\sigma}$ and $t > r_0$. The uniqueness of the propagator yields $\CR{\tilde{U} } (t,r_0) = U(t, r_0)$.

\subsection{Density of domain of wave operators}
In the proofs of the theorems, we assume that the initial state $\phi $ is included in the space $ \Omega _{0}$, which is
$$ \Omega _{0} := \left\{ {\phi} \in \SCR{S} ({\bf R}^n ) \, \Big| \,  \hat{\phi} \in C_0^{\infty} ({\bf R}^n \backslash \{ 0 \}) \right\}. $$ However, focusing on \eqref{j13}, $\phi$ must be written as the form $\phi = U_0(r_0,0) \psi $ for some $\psi \in \CR{ \SCR{S} ({\bf R}^n) }$. Hence, we prove the following lemma: 
\begin{lem}\label{L1}
Let us define a space $\hat{C}_{0}$ as 
\begin{align*}
& \Big\{ \psi \in \SCR{S} ({\bf R}^n) \, | \,  (\SCR{F}[\SCR{U_0}(r_0,0){\psi}] ) \in C_0^{\infty} ({\bf R}^n \backslash \{ 0 \}) \Big\} \CR{.} 
\end{align*}
Then, $\hat{C}_0 $ is dense on $L^2({\bf R}^n)$.
\end{lem}
\Proof{ 
First, we assume that $\zeta _2 '(r_0) \neq 0$ and define a space 
\begin{align*}
B_0 := \left\{ 
\phi \in \SCR{S} ({\bf R}^n) \, | \, \SCR{F}[e^{im \zeta _1 '(r_0) x^2 /(2 \zeta _2 '(r_0))} \phi ] \in C_0^{\infty} ({\bf R}^n \backslash \{ 0 \})
\right\}.
\end{align*}
Therefore, $B_0$ is dense on $L^2({\bf R}^n)$, i.e., for all $\tilde{\ep} >0$ and for all $\psi \in L^2({\bf R}^n)$, there exists $\Phi \in B_0 $ such that 
\begin{align}
\left\| 
\psi - \Phi
\right\| \leq \tilde{\ep} \label{j11}
\end{align}
holds. By the definition of $B_0$, there exists $0< \ep < R$ such that 
\begin{align}
& \nn e^{im \zeta _1 '(r_0) x^2 /(2 \zeta _2 '(r_0))} \Phi = \chi_0 (\ep \leq p^2 \leq R) e^{im \zeta _1 '(r_0) x^2 /(2 \zeta _2 '(r_0))} \Phi \\ 
& \nn \Rightarrow 
\Phi = e^{-im \zeta _1 '(r_0) x^2 /(2 \zeta _2 '(r_0))}   \chi_0 (\ep \leq p^2 \leq R) e^{im \zeta _1 '(r_0) x^2 /(2 \zeta _2 '(r_0))} \Phi \\ 
& \nn \Rightarrow 
\Phi =   \chi_0 (\ep \leq (p + m \zeta _1 '(r_0) x/ \zeta _2 '(r_0) ) ^2  \leq R) \Phi \\ 
& \label{as1} \Rightarrow \Phi = 
 \chi_0 (\ep (\zeta _2 '(r_0)) ^2   \leq (  \zeta _2 '(r_0)  p + m \zeta _1 '(r_0) x )^2 \leq R (\zeta _2 ' (r_0)) ^2 ) \Phi 
 \end{align}
where $\chi _0 \in C_0^{\infty} ({\bf R})$ is a smooth cut-off function such that $\chi_0 (\ep \leq t \leq R) \equiv 1$ if $\ep \leq \CR{t} \leq R$ and $\equiv 0$ if $0 \leq \CR{t} \leq \ep /2$ or $2R \leq \CR{t} $. 
Because the classical trajectory satisfies \CR{$p_0 (t) = \zeta _2 '(t)p + m \zeta _1 '(t) x $} on $C_0^{\infty}({\bf R}^n)$, for $u \in C_0^{\infty} ({\bf R}^n)$, we have  
\begin{align*} 
\CR{ U_0 (r_0, 0)^{\ast} p^2 U_0(r_0,0) u = ( \zeta _2 '(r_0)  p + m \zeta _1 '(r_0) x)^2 u. } 
\end{align*} 
Then, the equality \eqref{as1} is equivalent to 
\begin{align} \label{j10}
  \Phi = 
U_0 (r_0,0)^{\ast} \chi_0 (\ep (\zeta _2 '(r_0)) ^2   \leq p^2 \leq R (\zeta _2 ' (r_0)) ^2 ) U_0(r_0,0) \Phi
\end{align}
By \eqref{j11} and \eqref{j10}, we obtain
\begin{align} \label{j12}
 \left\| 
U_0 (r_0 ,0) \psi - \chi_0 (\ep (\zeta _2 '(r_0)) ^2   \leq p^2 \leq R (\zeta _2 ' (r_0)) ^2 ) U_0(r_0,0) \Phi
\right\| \leq \tilde{\ep}. 
\end{align}
As $\chi _0 \in C_0^{\infty} ({\bf R}^n)$, we found that $\chi_0 (\ep (\zeta _2 '(r_0)) ^2   \leq p^2 \leq R (\zeta _2 ' (r_0)) ^2 ) U_0(r_0,0) \Phi \in \hat{C}_0$. Together with the unitarity of $U_0 (r_0,0)$, \eqref{j12} implies that for all $\Psi \in L^2({\bf R}^n)$ and $\tilde{\ep} > 0$, there exists $f  \in \hat{C}_0$ such that  
\begin{align*}
\left\| 
\Psi - f 
\right\|  \leq \tilde{\ep}. 
\end{align*}
Hence, $\hat{C}_0$ is dense on $L^2({\bf R}^n)$ when $\zeta _2 '(r _0) \neq 0$. 

Next, we consider the case where $\zeta _2 '(r_0) = 0$ but $\zeta _1 '(r_0) \neq 0$. We set a space 
\begin{align*}
A_0 := \left\{ 
\phi \in \SCR{S} ({\bf R}^n) \, | \,  e^{-i \zeta _2 '(r_0)p^2 /(2m \zeta _1 '(r_0)) } \phi  \in C_0^{\infty} ({\bf R}^n \backslash \{ 0 \}) \right\}. 
\end{align*}
Therefore, $A_0$ is dense on $L^2({\bf R}^n)$ as $e^{-i \zeta _2 '(r_0)p^2 /(2m \zeta _1 '(r_0)) } \phi = \phi $ holds when $\zeta _2'(r_0) =0$. Additionally, by the argument above, for all $\tilde{\ep} >0$ and $\psi \in L^2({\bf R}^n)$, there exists $\tilde{\Phi} \in A_0$ such that 
\begin{align*}
\left\| 
\psi - \tilde{\Phi}
\right\|  \leq \tilde{\ep}.
\end{align*}
By the definition of $A_0$, there exists $0< \ep <R$ such that 
\begin{align*}
&e^{-i \zeta _2 '(r_0)p^2 /(2m \zeta _1 '(r_0))}  \tilde{\Phi} = \chi_0 (\ep \leq x^2 \leq R) e^{-i \zeta _2 '(r_0)p^2 /(2m \zeta _1 '(r_0)) }\tilde{\Phi} \\ &
\Rightarrow 
\tilde{\Phi} =  \chi_0 (\ep (m \zeta _1 '(r_0)^2 \leq \left( m \zeta _1 '(r_0) x + \zeta _2 '(r_0) p \right)^2 \leq R   (m \zeta _1 '(r_0)^2) )  \tilde{\Phi} \\ & \Rightarrow 
\tilde{\Phi} = U_0(r_0,0)^{\ast} \chi_0 (\ep (m \zeta _1 '(r_0)^2 \leq p^2 \leq R   (m \zeta _1 '(r_0)^2) )  U_0(r_0,0)\tilde{\Phi}. 
\end{align*}
Hence, by the argument in the case where $\zeta _2 '(r_0) \neq 0$, we found that $\hat{C}_0$ is dense on $L^2({\bf R}^n)$. 

Hence, a problem occurs only when $\zeta _1 '(r_0)=\zeta _2'(r_0) =0$. However, the fundamental solutions satisfy 
\begin{align*}
\zeta _1 (t) \zeta _2 '(t) - \zeta _1 '(t) \zeta _2 (t) = 1 , \quad \mbox{for all } t \in {\bf R}, 
\end{align*} 
in which at least either $\zeta _1'(r_0)$ or $\zeta _2'(r_0)$ will not be $0$, thus completing the proof.
}

~~ \\ ~~ \\ 
{\bf Acknowledgments.} The first author is partially supported by the Grant-in-Aid for Young Scientists (B) \# 16K17633 from JSPS.


\begin{thebibliography}{99}
\bibitem{Do} Dollard, J.D.: Quantum-mechanical scattering theory for short-range and Coulomb interactions, Rocky Mountain J. Math. \textbf{1} 5--81, (1971). 

\bibitem{F} Fujiwara, D.: Construction of the fundamental solution for the Schr\"{o}dinger equation, J. D'Anal Math. \textbf{35}, 42 (1979). 

\bibitem{GMT} Geluk, J. L., Mari\'{c}, V., Tomi\'{c}, M.: On regularly varying solutions of second
order linear differential equations, Differential and Integral Eqn.
\textbf{6}, 329 (1993).

\bibitem{I} Ishida, A.: The borderline of the short-range condition for the repulsive Hamiltonian, J. Math. Anal. Appl. \textbf{438}, 267 (2016). 

\bibitem{I2} Ishida, A.: Nonexistence of usual wave operators for fractional Laplacian
and slowly decaying potentials, East Asian J. Appl. Math. \textbf{9}, 233 (2019).

\bibitem{JO} Jensen, A., Ozawa, T.: Existence and non-existence results for wave operators of the Laplacian, Rev. Math. Phys., \textbf{5}, 601-629, (1993).  

\bibitem{Ka} Kawamoto, M.: Quantum scattering for time-decaying harmonic oscillators, preprint, arXiv:1704.03714  

\bibitem{Ko} Korotyaev, E. L.: On scattering in an external,
	    homogeneous, time-periodic magnetic field,
	    Math. USSR-Sb., \textbf{66}, 499-522, (1990).

\bibitem{Na} Naito, M.: Asymptotic behavior of solutions of second
	    order differential equations with integrable coefficients,
	    Trans. A.M.S., \textbf{282}, 577-588, (1984). 

\bibitem{O} Ozawa, T.: Non-existence of wave operators for Stark effect Hamiltonians, Math. Z., \textbf{207}, 335-339, (1991). 
 
\bibitem{Ya} Yajima, K.: Schr\"{o}dinger evolution equation with magnetic fields, J. Analyse. Math., \textbf{56}, 29--76 (1991). 
 
 
\end{thebibliography}
\end{document}